\documentclass[aps,prl,twocolumn]{revtex4}
\usepackage{epsfig}

\newcommand{\beqa}{\begin{eqnarray}}
\newcommand{\eeqa}{\end{eqnarray}}
\newcommand{\beq}{\begin{equation}}
\newcommand{\eeq}{\end{equation}}

\def\Journal#1#2#3#4{{#1} {\bf #2}, #3 (#4)}

\def\mod {\mbox{~mod}}
\def\m{{\tt m}}

\def\v{{\tt v}}
\def\f{{\tt f}}

\def\Ns{N_\phi}
\def\L{{\ell_{_H}}}
\def\LL{{\ell_{_H}}^2}

\def\T{\mathcal{T}}
\def\z0j{z_{0j}}
\def\z0k{z_{0k}}
\def\8{\infty}

\def\z0k{z_{0k}}

\def\Psiz0{\Psi_{(z_{0})}}

\begin{document}
\textsl{}
\title {Quasi-particle Tunneling Through a Barrier in the Fractional Quantum Hall Regime}
\author{Elad Shopen$^1$,
Yuval Gefen$^2$, Yigal Meir$^{1,3}$}
\address{$^1$Department of Physics, Ben-Gurion University, Beer-Sheva 84105, Israel}
%{\rm E-mail:} {\tt felad@bgu.ac.il, ymeir@bgu.ac.il } }
\address{$^2$Department of Condensed Matter Physics, The Weizmann Institute of Science, Rehovot 76100, Israel}
\address{$^3$The Ilse Katz Center for Meso- and Nano-Scale Science and  Technology, Ben-Gurion University,
Beer-Sheva 84105, Israel}

\begin{abstract}
Tunneling of fractionally charged quasi-particles (QPs) \emph{
through a barrier} is considered in the context of a multiply
connected geometry. In this geometry global constraints do not
prohibit such a tunneling process. The tunneling amplitude is
evaluated and the crossover from mesoscopic QP-dominated to
electron-dominated tunneling as the system's size is increased is
found. The presence of disorder enhances both electron and QP
tunneling rates.
\end{abstract}

\maketitle

One of the most remarkable facts about the fractional quantum Hall
effect (FQHE) is the existence of fractionally charged
quasi-particles (QPs). Their dynamics is manifest in a host of
physical phenomena, whose observation strongly supported the
veracity of Laughlin's theory\cite{laughlin}. It has been pointed
out \cite{QPversEL, gefenless} that QP tunneling is distinctly
different from electron tunneling. Perturbative
renormalization-group analysis \cite{kanefish} has indicated that in
the weak backscattering limit inter-edge tunneling through the FQHE
liquid is dominated by QP tunneling. These predictions have been
confirmed by experiments \cite{experiments}. In the opposite limit
of strong backscattering (nearly disconnected FQHE systems coupled
by weak tunneling through an insulator), the same RG analysis would
have predicted that tunneling should be dominated again by QP
tunneling. Common wisdom, however, has it that in this limit only
electron tunneling is possible. The rationale for that goes as
follows: consider two FQHE puddles weakly connected through
tunneling. The total number of electrons on each puddle is (nearly)
a good quantum number; hence it must be an integer. QP tunneling
would render this number non-integer, therefore such a process must
be excluded.

Our starting point here is to note that there are setups where the
above mentioned "global constraint" (i.e. the number of electrons on
each side of the barrier being an integer) does not exclude a-priori
QP tunneling through a potential barrier. The common wisdom alluded
to above needs then to be re-examined. Studying these setups is
particulary interesting in view of recent experimental results
\cite{recent experiments} which suggest the coexistence of both
electron and QP tunneling under strong backscattering conditions.

Consider first the annulus depicted in Fig.~1(a). Clearly, the
passage of a QP through the barrier would not violate the 'global
constraint'. There are two possible trajectories (for the presumed
noiseless incoming current) to traverse the system: either following
the edge adiabatically, or by tunneling through the barrier. The
outgoing current would then be noisy \cite{note:noise at low
frequency}. By analyzing this non-equilibrium noise one may detect
the effective charge involved.

Our extensive analysis, performed on a torus geometry described
below, leads to three main results: \textbf{(i)} For our
multiply-connected geometry and in the presence of a real potential
barrier we confirm the existence of QP tunneling, which decreases
rapidly with system size (Eq.~\ref{eq:torus}). \textbf{(ii)} We
study the amplitude of such QP tunneling processes and identify the
crossover, in terms of the system's parameters, between the
electron-tunneling-dominated and the QP-dominated regimes.
\textbf{(iii)} We show that in FQHE systems, the presence of
impurities may enhance both electron and QP tunneling amplitudes
(Eq.~\ref{eq:impurities}), in the spirit of the
Shklovskiii-Li-Thouless mechanism \cite{lithouless}.

\begin{figure}[h]
\epsfxsize=3.5in
\begin{center}
\leavevmode \epsffile{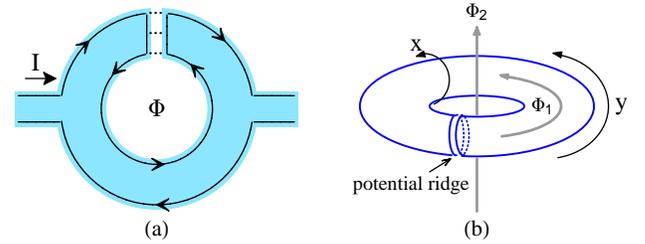}
\end{center}
\vskip -0.5truecm \caption{\footnotesize{(a) Proposed experimental
setup for an annulus, allowing the tunneling of both QPs and
electrons through a barrier. The edge states are marked. (b) The
torus geometry studied in this work; x and y represent two Cartesian
coordinates, the unit cell being $2\pi R \times L$; $\Phi_1$ and
$\Phi_2$ are two Aharonov-Bohm fluxes threading the torus. A Hall
liquid (of density $1/m$) covers the torus everywhere, except around
the barrier which is initially dry ("extended-holes"). Here the
barrier potential is circularly symmetric.}} \label{fig:per_geom}
\end{figure}

{\it The torus geometry}. To facilitate our analytical study and to
avoid complications emanating from the system being open, we
hereafter focus on a setup defined on a torus, spanned by the two
periodic coordinates ${0<x<2\pi R}$ and $0<y<L$ (Fig.~1(b)), with a
uniform magnetic field perpendicular to the surface of the torus. On
top of the torus surface we introduce a circular potential ridge
($V_0$). The tunneling investigated here is between the two sides of
this barrier. The main steps in the analysis are: \textbf{(i)} We
first construct modified Laughlin-Haldane-Rezayi states which
corresponds to a bulk filling factor $1/\m$ (the "wet area") and a
"dry area" (made of extended holes) where the electron density is
suppressed, ideally to zero. The ground state configuration is
obtained by maximizing the overlap of the dry area of the many-body
configuration with the barrier \cite{note:number_of_extended_holes}.
\textbf{(ii)} The Hamiltonian, hence the wave-functions, depend on
two gauge fluxes, $\Phi_1$ and $\Phi_2$ (Fig.~1(b)). By
adiabatically increasing $\Phi_1$ any many-body configuration will
slide rigidly in the y-direction, giving rise to a change in its
energy. As the levels of two different configurations cross, the
ground state of the system changes abruptly. Below we show that such
a change corresponds to a tunneling event. The set of many-body
energy levels is plotted in Fig.~2(a). \textbf{(iii)} To enable
tunneling we break the circular symmetry (in the x-direction),
introducing an additional asymmetric potential ($V_1$). This gives
rise to finite matrix elements between different configurations.
Avoided-crossing gaps in the energy-flux spectrum (Fig.~2(a)) are a
manifestation of tunneling: the period in flux reflects the nature
of the particle that tunnels. Below we calculate these tunneling
matrix elements. \textbf{(iv)} We demonstrate quantitatively how the
presence of a multitude of $\delta$-function impurities enhances the
tunneling.

\begin{figure}[h]
\epsfxsize=3.9in
\begin{center}\hskip -1.4truecm
\leavevmode \epsffile{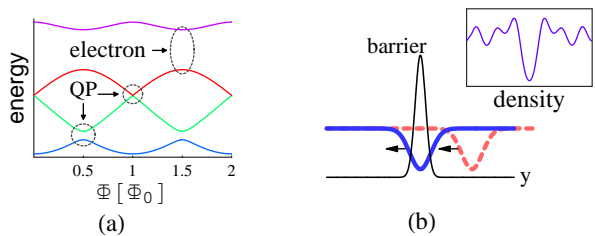}
\end{center}
%\SkpVert
\caption{\footnotesize{(a) Intersecting energy curves as a
function of the flux $\Phi_1$. Avoided crossings correspond to
tunneling of electrons and QPs as indicated. For QP (electron)
tunneling the periodicity of the adiabatically varied ground-state
energy is $\phi_0$ ($3 \phi_0$) \cite{gefenless}. (b) Schematic
density profile of the initial ground state $\Psi_0$ (solid line)
and the first excited state $\Psi_1$ (dashed line). As $\Phi_1$
increases, these density profiles slide to the left, $\Psi_1$
eventually becomes the ground state. (inset) An actual density
profile for ${\m=3, N=6, N_h=1}$. Densities corresponding to other
wavefunctions are rigid shifts thereof.}} \label{fig:the_tunneling
graph}
\end{figure}

Our initial Hamiltonian is $H=H_0+U_{int}+V_0$, where $H_0$
includes the kinetic part as well as the magnetic field and fluxes
and $U_{int}$ is the two-particle interaction. The barrier
potential $V_0$ is assumed to be sufficiently weak to exclude
mixing with higher Landau levels.

For $V_0=0$ and no dry area Haldane and Rezayi  \cite{haldane} have
found a set of $\m$ degenerate Laughlin wave functions. The solution
is obtained for a magnetic field which is quantized according to the
Dirac's condition $R L=\LL \Ns$, where $\Ns$ is the number of
magnetic flux quanta perpendicular to the torus surface and $\L
\equiv \sqrt{\hbar c/e B}$ is the magnetic length. These
wavefunctions are eigenfunctions of total quasi-momentum (TQM)
\cite{note:TQM} and, similarly to the Laughlin wave function
\cite{trugmankivelson}, are exact (zero energy) ground states of the
Hamiltonian with hard core interaction (i.e.
$\nabla^2\delta(\vec{r})$). Their ground state electron density is
nearly uniform. Here $\Ns=\m N$, where $N$ is the number of
electrons.

{\it Extended Hole Wavefunctions.} To render the barrier (and its
close vicinity) "dry", we tune the magnetic field to allow for $N_h$
holes: $\Ns=\m N+N_h$. The lowest Landau level consists of $\Ns$
single particle states $|n>$ (cf. Ref.\cite{haldane}). The density
profile of each single particle state is approximately a Gaussian in
the y-direction and uniform in the x-direction. The distance between
the guiding centers of adjacent states is $L/\Ns$. In the subspace
of the lowest Landau level (spanned by $({_{~N}^{~\Ns}})$ possible
Slater Determinants) the ground state is determined solely by the
interaction. Diagonalizing the hard-core interaction results in a
set of zero-energy ground states, each having $N_h$ holes extended
in the x-direction. As an example, consider $N_h=1$. When the
interaction term is diagonalized one obtains $\Ns$ Laughlin-like
states of zero energy. Each of these states corresponds to a
non-uniform electron density: the filling is $1/\m~$ almost
everywhere, but the occupation of one of the single-particle
quasi-momentum states $|n>$ is suppressed: the area around this
guiding center is "dry" (Fig.~2(b)). We denote such a many-body
state with an extended hole at $|n>$ by $\Psi_{n}$. It is an
eigenstate of the TQM. By sliding all guiding centers rigidly by $1$
we increase the TQM by $N$, shifting $\Psi_{n} \rightarrow
\Psi_{n+1}$. The above procedure is readily generalized to $N_h>1$.
Choosing the $N_h$ holes to be contiguous leads to a dry area of a
desired width \cite{note:extended hole depth}.

{\it tunneling: electrons versus QPs.} The initial ground state
whose dry area coincides with the barrier is denoted by $\Psi_0$. As
the flux $\Phi_1$ is varied adiabatically this many-body
configuration slides rigidly around the torus. The dry area of
$\Psi_0$ moves and its energy $E_0(\Phi_1)$ increases. The dry area
of $\Psi_{1}$, whose TQM differs from that of $\Psi_0$ by $N
\mod(\Ns)$, slides towards the barrier and its energy $E_1(\Phi_1)$
decreases. When $\Phi_1$ is increased by $\phi_0/2$, $E_0(\Phi_1)$
and $E_1(\Phi_1)$ intersect, and the ground state of the system
switches $\Psi_0 \rightarrow \Psi_1$. This corresponds to a shift of
each single electron state by $1$ : $|n> \rightarrow |{n+1}>$. Since
the average occupation of $|n>$ is $1/\m$ this process describes a
shift of charge of $e/\m$ from one side of the barrier to the other,
i.e. QP tunneling. The level-crossing degeneracy is lifted by
breaking the circular symmetry of the potential $V_0$: $H
\rightarrow H+V_1$. The QP tunneling matrix element is
$<\Psi_0|V_1|\Psi_{1}>$. By analogy, the two many-body states
$\Psi_0$ and $\Psi_{\m}$ differing in their TQM by $\m N \mod(\Ns)$
will cross when $\Phi_1$ increases by $\m \phi_0/2$, hence
$<\Psi_0|V_1|\Psi_{\m}>$ is the matrix element for electron
tunneling.
\\
In order to evaluate these tunneling matrix elements we use the fact
that $V_1$ is a single-particle potential. The overlap
$<\Psi_0|V_1|\Psi_p>$ ($p=1,\m$ for QP, electron tunneling
respectively) consists of contributions from the respective Slater
determinants components ${|k_1,\ldots,k_N> \in \Psi_0}$ and
${|\ell_1,\ldots,\ell_N> \in  \Psi_p}$ which are identical except
for a single pair $\tilde{k}, \tilde{\ell}$. The difference in TQM
is
\begin{equation}\label{TQMdifference}
\tilde{\ell}-\tilde{k}=pN \mbox{mod}(\Ns).
\end{equation}

Taking $V_1=\tilde{V}_1 \delta(x)$ renders the procedure
particularly simple \cite{note:V1 not delta(x)}. Then
$<\tilde{k}|V_1|\tilde{\ell}>\equiv \v_{p}$ depends only on the
difference $\tilde{\ell}-\tilde{k}$. The tunneling amplitude can
then be cast as $\T_p \equiv <\Psi_0|V_1|\Psi_p>=\f_{p}(L) \v_{p}$,
where $\f_p$ is a combinatorial factor resulting from summation over
all possible pairs satisfying (\ref{TQMdifference}), and is
calculated numerically. For $\v_{p}$ one readily obtains

\begin{equation}\label{v_p}
\v_p=\frac{\tilde{V}_1}{R} \sum_{q=-\8}^{\8} e^{-iq\Phi_2}
\exp{\left\{-\left[(p-q\Ns)\mbox{$\frac{L}{2 \Ns
\L}$}\right]^2\right\}}.
\end{equation}
\\
To leading order $\v_p\propto \exp[-(pN ~\mbox{mod} \Ns
~\mbox{$\frac{L}{2 \Ns\L}$})^2]$. This reflects the overlap of two
Gaussians separated by a distance $p N L/\Ns$ on the torus
(Fig.~3(a)). For a QP the separation is of the order $L/\m$ ($\Ns=\m
N+N_h$), so $\v_p$ scales as $\exp[-(L/2 \m \L)^2]$. By contrast,
for electrons the separation distance (defined modulo $L$) is
$N_hL/\Ns$, hence $\v_p$ scales as $\exp[-(L N_h /2 \Ns \L
)^2]=\exp[-(N_h \L/R)^2]$, which is L-\emph{independent} (where we
used the Dirac condition). Likewise we find that the factor $\f_p$
is roughly system-size independent for electron tunneling, while it
rapidly decreases (Gaussian-like) for QPs.

\begin{figure}[h]
\epsfxsize=3.6 in
\begin{center}
\leavevmode \epsffile{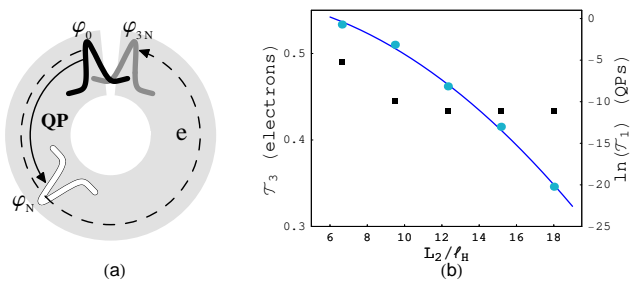}
\end{center}\vskip -0.5truecm
\caption{\footnotesize{Tunneling involves initial and final states
whose TQM differ by $pN$, $p=1,3$ for QPs and electrons
respectively. Here $\m=3$. The respective matrix elements involve
the overlap of two single-particle states (Gaussians), whose
separation is $pNL/\Ns$. For QPs the latter is of the order of
$L/\m$ (solid arrow, black and white Gaussians). Thus the overlap
factor for QPs, $v_{p=1}$, strongly decreases with $L$. By contrast,
for electrons (dashed arrow, black and grey) the distance (defined
modulo $L$) is $N_h L/\Ns << L$ and the dependence on $L$ is
negligible. (b) The calculated tunneling probabilities for electrons
($\T_3$) and QPs ($\T_1$), for $\m=3$, $N_h=1$. $\T_3$ is
practically independent of system size (squares), while for QPs
$\T_1 \sim e^{-\alpha L_2^2/ \LL}$ (circles). The solid line is a
fit with $\alpha \approx 0.07$.}} \label{fig:tunneling_numerics}
\end{figure}

We thus summarize
\begin{equation} \label{torus_results} \T_p \sim \left\{
\begin{array}{ll}
    e^{-\alpha L^2/ \LL}, & \mbox{QPs} ~(p=1)\\
    \mbox{$L$ independant}, & \hbox{electrons} ~(p=3). \\
\end{array}
\right. \label{eq:torus}
\end{equation}

For $N_h=1$ (and $N=2,\ldots, 6$) we obtain numerically $\alpha
\approx 0.07$ (Fig.~3(b)). We expect this to approach $1/12$ (${=1/4
\m}$) in the thermodynamic limit (cf. Eq.~(\ref{fixed number of
impurities}) for $N_{imp}=1$). Note that the factor $\f_{p=1}$
further suppresses the single-particle term $e^{-(L/2 \m \L)^2}
\rightarrow e^{-(L/2 \L)^2/\m}$.

{\it impurity-assisted tunneling}. We next introduce impurities into
the system, ${\cal H} = H_0+U_{int}+V_2$, where $V_2\equiv
\sum_{j=1}^{N_{imp}} V_{imp} \delta(z-z_{0j})$ represents $N_{imp}$
localized impurities ($V_{imp}>0$). Shklovskii and Li \& Thouless
\cite{lithouless} have shown that for non-interacting electrons,
impurities modify the Gaussian decay of the edge-to-edge Green
function into exponential. This result is generalized here to the
FQH regime. To simplify the analysis we employ the fact that QP
tunneling may be interpreted as a non-local process taking place
through the liquid, while electron tunneling takes place through
both the potential barrier and the liquid. Thus, the torus can be
effectively replaced by two cylinders whose circumference is $2 \pi
R$ and whose lengths are $L_{barrier} \equiv N_h L/\Ns$ and
$L_{liquid} \equiv L-L_{barrier} = \m N L/\Ns$. For a cylinder with
impurities the ground state wavefunction $\Psi_{\{{\bf z_{0}}\}}$
contains $N_{imp}$ localized holes at the impurity positions
\begin{eqnarray}\label{Psi}
\Psi_{\{{\bf z_{0}}\}}= \prod\limits_{j=1}^{N_h}
\prod\limits_{k=1}^N (e^{-i z_k/R}-e^{-i z_{0j}/R}) \Psi_L %\nn
%\label{eq:psiz0}
\end{eqnarray}
($\Psi_L=\prod_{i<j}(e^{-iz_i/R}-e^{-iz_j/R})^{\m}~ e^{-\sum_j
y_j^2/2}$ is Laughlin's cylinder wave function).

To obtain the various tunneling matrix elements we calculate the
overlap between $\Psi_{\{{\bf z_{0}}\}}$ and its shifted version
$\tilde\Psi_{\{{\bf z_{0}}\}}=\prod_{j=1}^N e^{- i p
z_j/R}\Psi_{\{{\bf z_{0}}\}}$, $p=1,\m$ for QP and electron
tunneling respectively. For $N_{imp}=1$ one recovers (up to
prefactors) the perturbative result \cite{assa}: QPs tunnel more
efficiently than electrons along a cylinder. We have evaluated this
overlap numerically for systems with $N \leq 6$ (for electrons
tunneling at integer filling we have considered $N \leq 17$) and
$N_{imp} \leq 4$, for \textbf{(i)} impurities equally spaced on a
line and \textbf{(ii)} at random positions throughout the cylinder,
averaging over $\sim1000$ realizations. For case \textbf{(i)} we
find that
\begin{equation} \label{fixed number of impurities}
<\Psi_{\{{\bf z_{0}}\}}|\tilde \Psi_{\{{\bf z_{0}}\}}> \sim \left\{
\begin{array}{ll}
    e^{-L^2/12 N_{imp} \LL}, &\mbox{QPs} \\%&\mbox{$\frac{e}{3}$ charged QP} \\
    e^{-L^2/4 N_{imp} \LL}, & \hbox{electrons}\\
\end{array}
\right.  \label{eq:impurities}
\end{equation}
agrees with the numerics. For case \textbf{(ii)} the decay factor of
the exponent is modified, but not the parametric dependence. We find
that when the longitudinal impurity density $\lambda=L/N_{imp}$ is
held constant, the typical hopping distance is kept fixed, a
Gaussian-to-exponential crossover takes place. This crossover can be
understood in terms of multiple impurity-assisted tunneling. For a
QP, as an example, $e^{-L^2/12 N_{imp}}\rightarrow e^{-\lambda
L/12}$.

{\it QP-electron crossover}. Studying this crossover is now
experimentally feasible \cite{crossover_experiments}. Here we
present a framework to study it theoretically in a multiply
connected geometry, e.g. the torus. As $L_{barrier}$ is varied
(compared with $L_{liquid}$) the QP tunneling $\T_1$ competes with
the electron tunneling $\T_3$. By calculating $\T_1$ and $\T_p$, as
explained above, varying the number of particles (modification of
$L_{liquid}$) and the number of extended holes (modification of
$L_{barrier}$), we obtain the ratio of the electron-tunneling
amplitude to that of QP tunneling (Fig. 4). This allows us to
determine the separation between the electron-tunneling dominated
and the QP-tunneling dominated regimes (dotted line). This can be
compared to the solid curve obtained by taking the estimates
$\T_1\sim e^{-L_{liquid}^2/4 \m \LL}$, $\T_3 \sim
e^{-L_{barrier}^2/4 \LL}+e^{-L_{liquid}^2/4 \LL}$.

\begin{figure}[h]
\epsfxsize=2.7in
\begin{center}
\leavevmode \epsffile{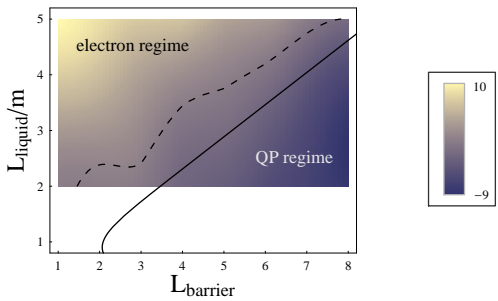}
\end{center}
\vskip -0.85truecm \caption{\footnotesize{Electron-QP crossover:
Plotted is the interpolation of $ln({\T}_3/{\T}_1)$ for discrete
values of $N$ and $N_h$ corresponding to $L_{liquid}/\m$ and
$L_{barrier}$ respectively (measured in units of $\LL/R$; here
$R=\sqrt{14 \pi}$). Dashed line: the crossover curve
${\T}_3={\T}_1$. Solid line: an approximated crossover-curve
obtained by taking $e^{-L_{liquid}^2/12 \LL} = e^{-L_{barrier}^2/4
\LL}+e^{-L_{liquid}^2/4 \LL}$.}} \label{fig:crossing}
\end{figure}

{\it discussion.} Keeping the barrier size fixed and increasing the
torus length we have found that the QP tunneling amplitude decreases
while the electron amplitude is mostly unaffected. This supports the
picture that the QP tunneling through a barrier \cite{helias}, while
in principle possible, is a mesoscopic effect. It may be interpreted
as a QP leaping through the liquid around the barrier. In the
thermodynamic limit the QP tunneling amplitude vanishes, in
accordance with common wisdom. The dependence of the tunneling
amplitude on length scales does not conform to the scaling resulting
from the RG treatment \cite{kanefish}. It is strongly modified by
the multiple connectedness of the system. Adding disorder enhances
the tunneling amplitudes. As can be seen from Eq.~(\ref{fixed number
of impurities}), special arrangements of the impurities can lead to
even stronger enhancement (e.g. increase the linear density of
$N_{imp}$, while keeping the two-dimensional density fixed). We
believe that the best candidate to test the ideas outlined here is
the annulus geometry (Fig.~1(a)). The relevance of the current
results to the annulus geometry will be explored in future studies.

{\it acknowledgments} We thank A. Altland, F.D.M. Haldane, B.
Halperin, B.I. Shklovskii and D.J. Thouless for illuminating
discussions on various aspects of the problem. This work was
supported by the US-Israel Binational Science Foundation, the Israel
Academy of Science and the Minerva Foundation. Y.G. was supported by
the AvH foundation. Y.M. was supported through the Einstein Minerva
Center for Theoretical Physics (through the BMBF).

\nopagebreak


\begin{thebibliography}{99}
\bibitem{note:noise at low frequency}
Note that only forward scattering is involved here. This means that
the transmission probability is $1$, implying absence of noise in
the d.c. limit. The noise alluded to above may be detected at finite
frequencies.
\bibitem{note:number_of_extended_holes}
For a ridge whose extension in the y-direction is $L_{barrier}$, the
number of extended holes required to render the ridge dry is
$N_h=[R~L_{barrier}/\LL] + 1$.
\bibitem{note:TQM}\
The total angular momentum is defined $\mod(\Ns)$.
\bibitem{note:analytic_extended_wf}
In the present papar we employ a numerical recipe for the
construction of extended hole states. We note that these may also be
obtained analytically, starting with Laughlin's state with localized
holes. As this requires some details, and is not crucial to
understanding the present analysis, we will present it elsewhere.
For $N_h=1$: $\Psi_n=\int g_n^{*}(z_0)~\Psi_{(z_0)} dz_{0}$ where
$g_n$'s are single particle states on the \emph{extended} unit cell
$2 \pi \m R \times \m L$, and $\Psi_{(z_0)}$ is a Laughlin state
with a localized hole at $z_0$ \cite{haldane}.
\bibitem{note:extended hole depth}
The width of a single-particle quasi-momentum state in the $y$
direction is $\L$; when $N_h < R/\L$ we have a reduced density
rather than a dry area.
\end{thebibliography}
\end{document}